\documentclass[aip,amsmath,amssymb,reprint]{revtex4-1}

\usepackage{amsmath}
\usepackage{graphicx}
\usepackage{dcolumn}
\usepackage{bm}
\usepackage{hyperref}
\usepackage[utf8]{inputenc}
\usepackage[T1]{fontenc}
\usepackage{mathptmx}
\usepackage{etoolbox}
\usepackage{subfig}
\usepackage{xcolor}
\usepackage{diffcoeff}
\usepackage{float}
\usepackage{afterpage}

\hypersetup{colorlinks=true,citecolor=blue,urlcolor=blue,linkcolor=blue}
\makeatletter
\def\@email#1#2{%
 \endgroup
 \patchcmd{\titleblock@produce}
  {\frontmatter@RRAPformat}
  {\frontmatter@RRAPformat{\produce@RRAP{*#1\href{mailto:#2}{#2}}}\frontmatter@RRAPformat}
  {}{}
}%
\makeatother
\begin{document}

\preprint{AIP/123-QED}

\title{Assessment of RANS simulation for hybrid ventilation in a reduced-scale classroom model}

\author{Deep Narayan Singh}
\author{Lagoon Biswal}
\affiliation{Department of Aerospace Engineering, Indian Institute of Technology Bombay, Mumbai 400076 India}
\author{Girish Naik}
\author{Manaswita Bose}
\affiliation{Department of Energy Science and Engineering, Indian Institute of Technology Bombay, Mumbai 400076 India}
\author{Krishnendu Sinha}
\email{krish@aero.iitb.ac.in}
\affiliation{Department of Aerospace Engineering, Indian Institute of Technology Bombay Mumbai 400076 India}
\date{\today}

\begin{abstract}

In this paper, we study the ventilation airflow in a model classroom, where exhaust fans throw out the used air, to replace it with outdoor air through open door. Hybrid ventilation, or mechanically assisted natural ventilation, of this kind is used as a retrofit design to reduce infection risk from airborne transmission.
The air stream entering the door forms a jet-like flow, driven by the suction effect of exhaust fans.
We compute the jet velocity using Reynolds averaged Navier Stokes (RANS) method and compare with velocity field measured using particle image velocimetry.
Different turbulence models are found to match experimental data near the door, but they over-predict the peak jet velocity further downstream.
There is minimal variation between the results obtained using different turbulence models.
%
The CFD results are found to be sensitive to inlet boundary conditions, whether the door entry is specified as a pressure inlet or velocity inlet. The geometry of the space outside the door also has a significant effect on the jet velocity.
Changing the boundary condition takes the CFD results closer to the experimental data; the velocity profiles computed with the extended domain being the closest to the measured peak velocity.
%
Interestingly, the centerline velocity decay computed with the extended domain aligns well with the experimental data. The other cases, irrespective of turbulence model, show much lower decay rate that seem to align with wall jet scaling.
This suggests that geometry and boundary conditions at the door is critical to predict the airflow in hybrid ventilation.


\end{abstract}

\maketitle


\section{\label{sec:level1} INTRODUCTION }

Indoor environments are critical to our health and well being, as we spend a significant amount of time daily in indoor spaces.
A good ventilation system in indoor spaces is essential for indoor air quality,\cite{zhang2021disease,Sha2023,Rencken2021} which plays a crucial role in human health.\cite{Foster2021,he2021airborne,Pandey2023}
This is especially true for high-occupancy spaces like classrooms, which are prone to air-borne infection spread between occupants.
Ventilation of indoor spaces can be achieved by mechanical systems, natural means,\cite{durrani2015evaluation, bhagat2020effects, Omrani2021} or mixed mode.\cite{industrial}

\if{false}
{\bf Ventilation} of indoor spaces is achieved either by mechanical or natural mode.
Mechanical ventilation systems used fans and ducts to provide "clean" air into an indoor space, while removing and recirculating the used air, with adequate filtration and possible disinfection.
On the other hand, natural ventilation is usually through open doors, windows, etc. -- either wind-driven or due to bouyancy, where outdoor cooler air is used to remove the heat from indoor spaces~\cite{durrani2015evaluation}.
In the absence of mechanically driven components, natural ventilation is far more energy efficient.
It is especially attractive for infection control, as the outdoor air is devoid of occupant generated bio-aerosol.
However, the effectiveness of natural ventilation depends on outdoor conditions, as well as the geometry of the building, etc.\cite{bhagat2020effects}
\fi 

Mixed mode or hybrid ventilation is also called mechanically-assisted natural ventilation, where fans are used to draw in natural air into a room. 
Mechanical exhausts are often used for removing gaseous contaminants in industrial setting.\cite{industrial}
Several researchers have studied the design of exhaust hood,\cite{hood} and exhaust speed to increase the efficiency of hybrid ventilation.\cite{exh_speed}
Exhaust fans or extractors are routinely used in kitchens and washrooms in residential setting. Mechanically-assisted natural ventilation via exhaust fans are also explored in hospital setting to convert existing ward to a temporary isolation room.\cite{isolation}
In this work, we consider the hybrid ventilation of a university classroom, as described below.

Design of an efficient, cost effective ventilation system requires an understanding of the airflow. Numerical simulations based on Computational Fluid Dynamics (CFD) are widely used to determine the velocity field in different configurations of enclosed spaces \cite{susin2009evaluating, nielsen2015fifty, zhang2024effects}, 
to estimate pollutant dispersion,\cite{Luo2023,Kumar2023,armand20253d} and to understand aerosol transport.\cite{abuhegazy2020numerical} CFD based simulations are often used as the base to propose new designs and operation strategies \cite{auvinen2022high, ritos2024effects, wang2025energy} of ventilation systems. 
The challenges in large-scale CFD simulations for non-standard flow geometries 
encountered in ventilation studies are multi-faceted: (a) selection of turbulence model; (b) the validation of the CFD results against realistic experiments;\cite{zhang2007evaluation, van2017accuracy} (c) the uncertainties in the flow boundary conditions.\cite{vanHoof2022}

\if{false}
{\bf Computational fluid dynamics} is routinely used to study and predict ventilation airflow,\cite{susin2009evaluating,nielsen2015fifty} 
pollutant dispersion~\cite{Luo2023,Kumar2023,armand20253d} and aerosol transport~\cite{abuhegazy2020numerical} in enclosed spaces. In particular, simulations based on Reynolds-averaged Navier Stokes (RANS) method are widely used for their cost competitiveness.\cite{CFD_review_zhai2007evaluation,blocken2018over}
Different geometries are modeled under varying ventilation scenarios, to evaluate the effect of different levels of occupancy and interventions.
CFD is also used to design indoor spaces and ventilation strategies~\cite{zhang2024effects} for effective use of resources to maintain comfortable and healthy indoor environment,~\cite{auvinen2022high,ritos2024effects} at minimal energy cost.~\cite{wang2025energy}
Validation of CFD simulations and methodologies with experimental data is a crucial part of ventilation studies. Specifically, turbulence models used in RANS methods are prone to modeling error which can limit the accuracy of the CFD solution.~\cite{van2017accuracy,zhang2007evaluation}
The effect of inlet boundary conditions have also been studied for generic forced ventilation scenario.\cite{inletBC}

A wide range of {\bf turbulence models} are available in the RANS framework. 
Commonly used models include eddy viscosity based models like k-epsilon,~\cite{lanuder1974numerical} k-omega,~\cite{menter1994two} Spalar Almaras,~\cite{SA_model} v2f~\cite{durbin1995separated} and their variants.
The accuracy of turbulence models is highly dependent in the flow physics. Some models are more accurate for wall-bounded flows,~\cite{patel1985turbulence} while others are tailored for free shear flows.~\cite{lanuder1974numerical} Rotational flows and those with adverse pressure gradient~\cite{SSTmenter1994two} are particularly challenging. Ventilation airflow are usually complex and involve a combination of different flow physics. It is therefore critical to test the validity of RANS turbulence models for a given ventilation scenario. A large amount of research work has been dedicated to evaluating the accuracy of turbulence models and validating RANS methods for indoor ventilation.

The CFD validation studies in literature are based on comparison with experimental data, either in full-scale rooms~\cite{hu2022comprehensive,villagran2019transient,caciolo2012numerical} or reduced-scale models.~\cite{ramponi2012cfd,ramponi2012cross-ventilation} Typical quantities of interest include the velocity vector field, pressure coefficient, velocity profiles, turbulent kinetic energy and Reynolds stresses. Experimental data is available for generic configurations representing natural convection in a tall cavity,~\cite{betts2000experiments} forced convection in a room with partitions,~\cite{ito2000model} mixed convection in a square cavity;~\cite{blay1992confined} and strong buoyant flow in a fire room.~\cite{murakami1995measurement} Different types of building geometry have been tested for cross-ventilation in wind-tunnel experiments.~\cite{KARAVA2011266}
The effect of inlet boundary conditions have also been studied for generic forced ventilation scenario.\cite{inletBC}
\fi 

It is hard to find consensus on which turbulence model to use in indoor ventilation. However, there is general agreement that mean flow characteristics are usually well-predicted by RANS models, while turbulence statistics are not reproduced correctly by RANS-based simulations. 
Flow unsteadiness is a key aspect in natural ventilation that is often missed by steady RANS simulation.
Large Eddy Simulation (LES) gives better accuracy, but it is not feasible for practical applications involving parametric variations and design studies. This is because of the higher computational cost up to 30 to 80 times RANS simulations.~\cite{sudirman2025validation,caciolo2012numerical}


Detailed comparison of the air velocities obtained from CFD simulations with PIV (Particle Image Velocimetry) based measurements is typically performed on reduced-scale geometries \cite{ramponi2012cross-ventilation, hu2022comprehensive, KARAVA2011266}. Validation of the numerical results against point measurement of temperature or velocity using anemometers is performed in full-scale rooms \cite{villagran2019transient, caciolo2012numerical}. Mechanical ventilation through duct flow is well studied, and the experiments by Nielsen have been used as a benchmark case for the last sixty years.\cite{Nielsen2010Annex} 
Natural or cross-ventilation experiments of generic configurations are commonly performed in wind tunnels \cite{ramponi2012cross-ventilation,KARAVA2011266, vanHoof2022}. 
%
%
Typical quantities of interest include the velocity vector field, pressure coefficient, velocity profiles, turbulent kinetic energy and Reynolds stresses. 
The effect of inlet boundary conditions have also been studied.\cite{vanHoof2022}
%

Detailed review of literature suggests that the majority of the validation studies focus on mechanical \citep{nielsen2015fifty, Nielsen2010Annex, ito2000model} or natural ventilation.\cite{vanHoof2022} There are very limited studies of hybrid ventilation scenarios. 
For mechanical ventilation, RANS turbulence models are usually found to match the mean velocity profile in the inlet jets (primary flow) driving the indoor circulation. There are however discrepancies in the secondary flow (recirculation or separated) regions. The models are also not able to predict the second-order quantities like Reynolds stresses and turbulent kinetic energy.
In case of natural ventilation, the comparison between RANS models and experiments is less promising. A recent paper reports 50-60\% agreement between mean velocity prediction and measurements (within experimental uncertainty) for a naturally ventilated reduced-scale model in a wind tunnel.\cite{RANS_accuracy_sudirman2025validation} 
This could be because of additional features in natural ventilation, like large scale  unsteadiness in the incoming jet \cite{vanHoof2022} and vena contracta.\cite{KARAVA2011266}

Hybrid ventilation literature mostly focuses on different application scenarios, using either field \cite{isolation} or laboratory experiments \cite{industrial} or
computational simulations.\cite{organic}
CFD validation is either with canonical test cases \cite{organic} or limited  data from hybrid ventilation experiments.\cite{industrial}
Careful validation with detailed experimental measurements is virtually non-existent to the best of our knowledge. This paper is an attempt to fill this research gap.
We consider a classroom scenario with hybrid ventilation
where the air circulation is driven by exhaust fans.
A closely coupled experimental and computational study is performed to compare the velocity field in the scaled-down breathing plane.
We focus on the inlet jet from the open door) that has not been studied much earlier for natural \cite{van2017accuracy} and hybrid ventilation.
We compare the velocity profiles at different streamwise locations and study the decay of the jet velocity with distance.
Different turbulence models are tested in CFD and the effect of different
boundary conditions at the inlet is investigated.

\if{false}
\textcolor{red}{
Detailed review of literature suggests that the majority of the studies focus on mechanically ventilated spaces \citep{nielsen2015fifty, Nielsen2010Annex, ito2000model}, or natural ventilation scenarios.\cite{vanHoof2022} There are very limited studies of hybrid ventilation scenarios.\cite{Hajdukiewiczetal2024}. 
Careful validation with experimental results is virtually non-existent, to the best of our knowledge. 
This paper is an attempt to fill this research gap. 
Exhaust fan driven mixed ventilation is common in warm environment. In this work, a naturally ventilated room with one open-door is considered in which the air circulation is driven by two exhaust fans. Selection of the geometry is motivated by one of the real scale classroom in one of the premier Indian institute. The specific objective is to  
compare the mean flow velocity predicted by CFD with those measured in experiments. Different turbulence models are tested and the effect of different boundary conditions is investigated. Results are also compared against the standard flow configurations \cite{Nielsen2010Annex} for comprehensive understanding. 
{\bf Need to enhance this part}}

The geometry of the classroom represents a simplified (cross ventilation) scenario with air entering through the door on one side, flowing along the front wall and exiting through the exhaust fans on the opposite wall.
The air circulation set up in the room is similar to that of forced convection studied extensively by Nielsen et al.\cite{Nielsen2010Annex} and Ito et al.,\cite{ito2000model} where the incoming stream forms a wall jet, with a large recirculation region adjacent to it.
Although qualitatively similar, we find some crucial differences between the mechanically ventilated configuration\cite{Nielsen2010Annex,ito2000model} and the current case of hybrid ventilation.
We study these differences in terms of the decay of the mean centerline velocity with distance, and compare the experimental and CFD results to well-established scaling relations in literature.


Nielsen's test cases have been extensively used for CFD validation. It is found that CFD predictions match experiments in the wall jet flow, while there are discrepancies in the recirculation region.
We tested several turbulence models for Nielsen's geometry and the results are consistent with those in literature.
The current scenario, on the other hand, poses significant \textbf{challenge for CFD}.
Although the broad features of the flow are captured, there are crucial differences, especially near the exhaust fans.
We present a detailed analysis of the velocity field obtained from CFD, specifically in relation to the \textbf{boundary conditions}, turbulence models and three-dimensional effects, in order to explain the mis-match with experimental data.

Finally, we look at the centerline velocity decay of Nielsen's expt, the current case of exhaust driven flow. Two additional cases with a forced jet at the door location are also studied, both experimentally and by CFD. The veloicyt decay shows interesting trends, when compared with established scaling of round jets and wall jets. The scaling arguments give important insight about airflow driven by the exhaust fans, whcih is used to bring out the limitations of CFD in such cases of indoor ventilation. We propose remedies and improvements, as well as future research directions to understand this seemingly simple ventilation test case.

\fi 

The paper is organised as follows. Section II presents the simulation methodology and its validation against a well-studied case of forced convection. The reduced-scale classroom model is described in section III, along with the experimental details. The computational domain, boundary conditions and grid are also included. This is followed by Results in Section IV, which describes the velocity and turbulence field, comparison of different turbulence models with experimental data and the effect of boundary conditions. Section V presents the scaling of centerline velocity in CFD and experiments, as well as existing scaling relations for the respective flows. Conclusions are in Section VI.


\section{SIMULATION METHODOLOGY}

The Reynolds-Averaged Navier-Stokes equation and the continuity equation for incompressible fluid are solved using the software package Ansys FLUENT 2022 R2. Simulations are performed with different turbulence models, including the well-known two-equation models such as, Renormalized $k-\epsilon$, Realizable $k-\epsilon$, SST-$k-\omega$, transition SST-$k-\omega$, and the one equation Sapart-Allmaras model. Table ~\ref{tab:turbmodel} presents a comprehensive comparison of the models.

The governing equations are solved using the pressure-velocity coupling with a second-order upwind scheme for the discretization of pressure and convective terms of the momentum equation, as well as turbulence transport equations. We have used hybrid initialization, and the iterative convergence for steady state is obtained when the scaled residuals drop below $10^{-6}$.

\begin{table*}
    \centering
    \caption{Summary of widely used turbulence models}
    \begin{tabular}{clll}\\ \hline \hline
    
         Sr. No.&Turbulence Model  &Description    & Typical Application\\ \hline
         \\ 1. &Renormalized Group & Incorporates statistical techniques  &It performs better in rapidly \\
         & $k-\epsilon$ model & from the Navier–Stokes equations  &strained, swirling, and \\
         & & and introduces an additional term &low-Reynolds number \\
         & & in the \ensuremath{\varepsilon} equation to model &flows.~\cite{orszag1993renormalisation} \\
         & & smaller-scale turbulence. &\\ \\
        2.  & Realizable $k-\epsilon$ model & Improves upon the standard $k-\epsilon$  & Ensure physical realizability \\
         &  &model formulation by modifying   & of the normal stress components \\
         &  & the turbulent viscosity and  & and improve the accuracy \\
         
         & &\ensuremath{\varepsilon} transport equation. &of the prediction of complex shear,  \\
         & & & rotating, and recirculating flows.~\cite{shih1995new}\\ \\
         3. & Shear Stress Transport & Combines the \textit{k}-\ensuremath{\omega}   &  It is suitable for \\
         & (SST) \textit{k}-\ensuremath{\omega}& formulation near walls and & adverse pressure gradients and \\
         & & the \textit{k}-\ensuremath{\varepsilon} model in the free stream  & flow separation prediction in aerospace \\
         & &using a blending function.  &and turbomachinery applications.~\cite{menter1994two}\\
         & & &It includes a shear-stress limiter. \\ \\
         4. & Transition SST \textit{k}-\ensuremath{\omega}& It builds upon the \textit{k}-\ensuremath{\omega} SST model& It captures the laminar-to-turbulent \\
         & &  by adding equations for intermittency &  transition, essential for accurate skin friction  \\
         & &  and the transition momentum& and separation predictions in low-turbulence\\
         & &  thickness Reynolds number.&environments such as turbine blades.\\ \\
        5. & Spalart--Allmaras model& One-equation model; &Though limited in accuracy \\
           & &It solves a transport equation for a &for free shear flows,\\
           & & modified turbulent viscosity and& it is robust and ideal for attached \\
           & & is computationally efficient.& boundary layer simulations.~\cite{spalart1992one}\\
         & & & \\
         \hline \hline
    \end{tabular}
    \label{tab:turbmodel}
\end{table*}

\begin{figure*}[t!]  
{\includegraphics[width=0.8\textwidth]{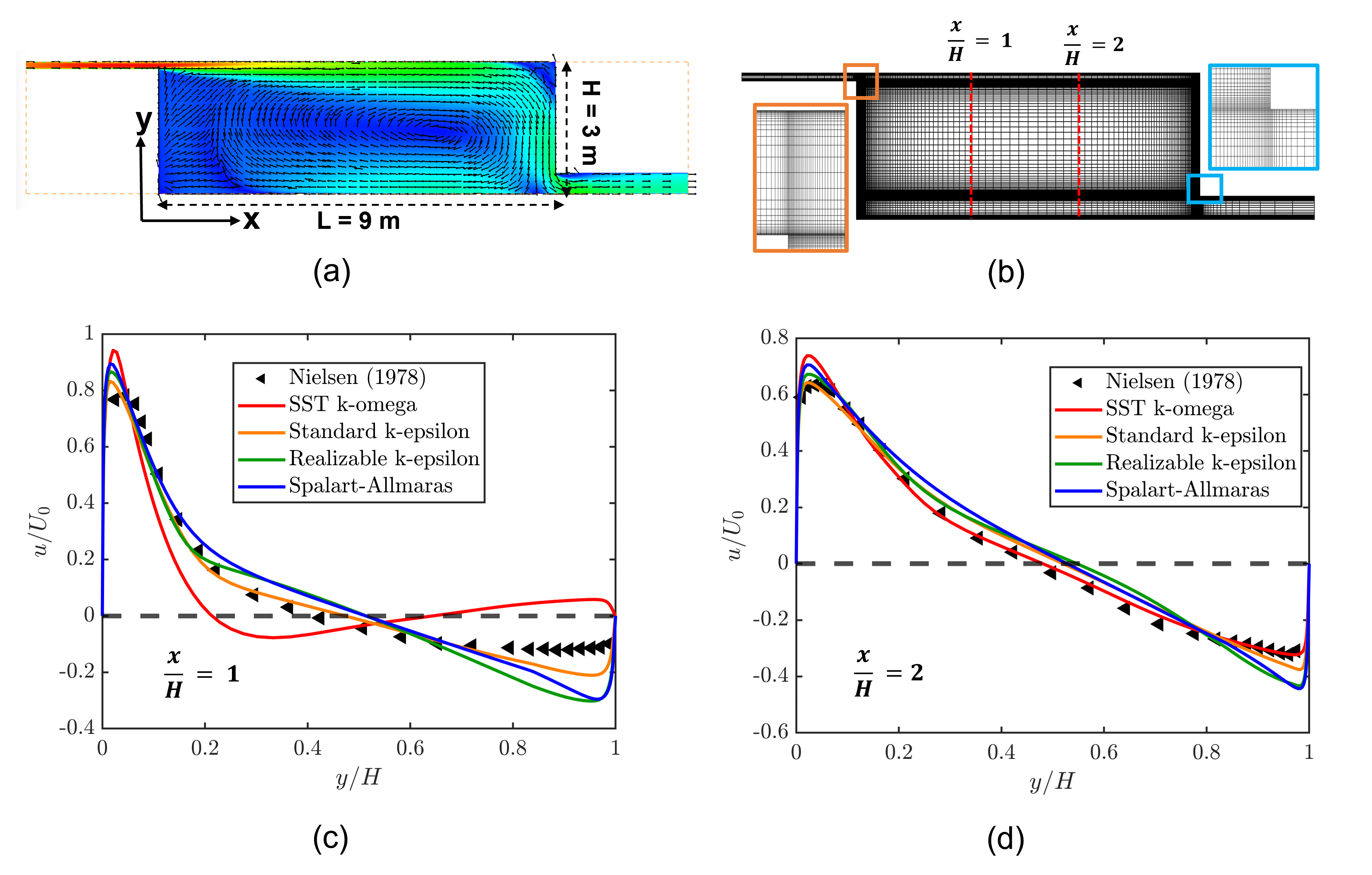}} 
\caption{Validation of CFD methodology with experimental data for forced convection case from Nielsen et al.\cite{Nielsen2010Annex} }
\label{fig:validation}
\end{figure*}

\if{false}

\subsection{Turbulence modelling}

We perform simulation with different turbulence models in order to validate with experimental results.  The turbulence models range from very well-known realizable $k-\epsilon$ model to lesser known recently developed transition models like transition SST. A brief description of the turbulence models is presented below. 
The majority of the results are generated using the realizable \textit{k}-\ensuremath{\varepsilon} model. Results generated using the other models are presented subsequently.

The realizable $k-\epsilon$ model improves upon the standard $k-\epsilon$ model formulation by modifying the turbulent viscosity and \ensuremath{\varepsilon} transport equation. Ensure physical realizability of the normal stress components and improve the accuracy of the prediction of complex shear, rotating, and recirculating flows.~\cite{shih1995new}



The Renormalization Group (RNG) \textit{k}-\ensuremath{\varepsilon} model incorporates statistical techniques from the Navier–Stokes equations and introduces an additional term in the \ensuremath{\varepsilon} equation to model smaller-scale turbulence. It performs better in rapidly strained, swirling, and low-Reynolds number flows.~\cite{orszag1993renormalisation}

The Shear Stress Transport (SST) \textit{k}-\ensuremath{\omega} model combines the \textit{k}-\ensuremath{\omega} formulation near walls and the \textit{k}-\ensuremath{\varepsilon} model in the free stream using a blending function. It includes a shear-stress limiter, making it suitable for adverse pressure gradients and flow separation prediction in aerospace and turbomachinery applications.~\cite{menter1994two}

The Transition SST model builds upon the SST \textit{k}-\ensuremath{\omega} model by adding equations for intermittency and the transition momentum thickness Reynolds number. It captures the laminar-to-turbulent transition, essential for accurate skin friction and separation predictions in low-turbulence environments such as turbine blades.

The Spalart--Allmaras model is a one-equation model tailored for wall-bounded aerospace flows. It solves a transport equation for a modified turbulent viscosity and is computationally efficient. Though limited in accuracy for free shear flows, it is robust and ideal for attached boundary layer simulations.~\cite{spalart1992one}

\fi

\subsection{Validation test case}

We consider the forced convection test case of Nielsen et al.\cite{nielsen1978velocity} that has been extensively used for CFD validation in the past. The geometry consists of an elongated room with duct entry on top left corner and exit from bottom right. The aspect ratio of the room is ($L/H = 3$, where $L$ is the length and $H$ is the vertical height. A schematic of the geometry, along with dimensions given in Fig.~\ref{fig:validation}a. The entry duct spans the entire width of the room, so that the flow is two dimensional in the majority of the domain. 
The forced jet from the inlet duct forms a wall jet along the ceiling. There is a large recirculation region in the rest of the room; see Fig.~\ref{fig:validation}a.
Velocity and turbulence measurements are reported at the spanwise mid-plane. These are used for CFD validation.

The inlet duct height is 0.168 m and inlet velocity is given as 0.455 m/s. The Reynolds number based on these quantities is 5000. The inlet flow is characterized in terms of mean velocity and streamwise Reynolds stress. These are used as boundary condition in the CFD simulations. 
Fig.~\ref{fig:validation}b shows the structured grid used in the simulation. It consist of $2.4$ million cells, with $y^+ \sim 1$ at the walls and refinement in the jet regions. This grid was found to be adequate based on a detailed refinement study.

The experimentally measured velocity profiles are plotted at two streamwise location $x/H$ = 1 and 2, as shown in Fig.~\ref{fig:validation}c and ~\ref{fig:validation}d. 
The high velocity in the region $y/H < 0.2$ corresponds to a wall jet, while relatively low velocities are obtained in the recirculation regions ($y/H > 0.2$). Negative velocities are reported near bottom wall (y/H $> 0.8$). This corresponds to the reversed flow in the recirculation region.

All the turbulence models, except SST $k-\omega$, reproduce the experimental data well at the first location ($x/H$ = 1); there are minor differences in the recirculation regions. The models match the qualitative trend, but underpredict the values for $y/H > 0.65$.  
On the other hand, SST $k-\omega$ gives very different results, with positive velocity near the bottom wall. This is opposite to the experimental data, and is caused by a secondary recirculation in this region. The trend is consistent with those presented in literature.~\cite{rong2008simulation}
At $x/H = 2$, all the turbulence models, match the experimental data fairly well. 

We note that the flow topology in the validation test case presented above is very similar to that observed in the present configuration. 
This includes a high velocity jet (primary flow) along a wall and a large recirculation zone (secondary flow) in the remainder of the room. 
The validation case therefore serves as a benchmark for comparison with the current data. 
The geometric configuration, along with the experimental and computational details, of the present case is described below.

$ $
\clearpage

\section{Reduced scale classroom model}

\begin{figure}[t!]  
\subfloat []{\includegraphics[width=0.4\textwidth]{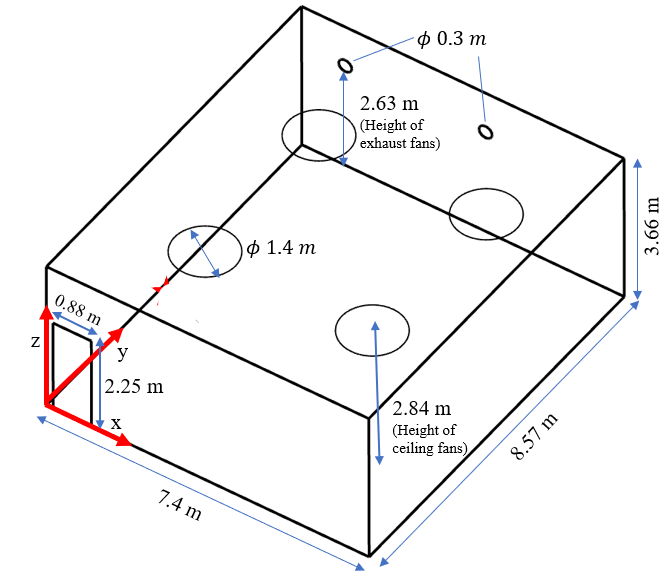} \label{fig:geom1}} \\
\subfloat [] {\includegraphics[width=0.45\textwidth]{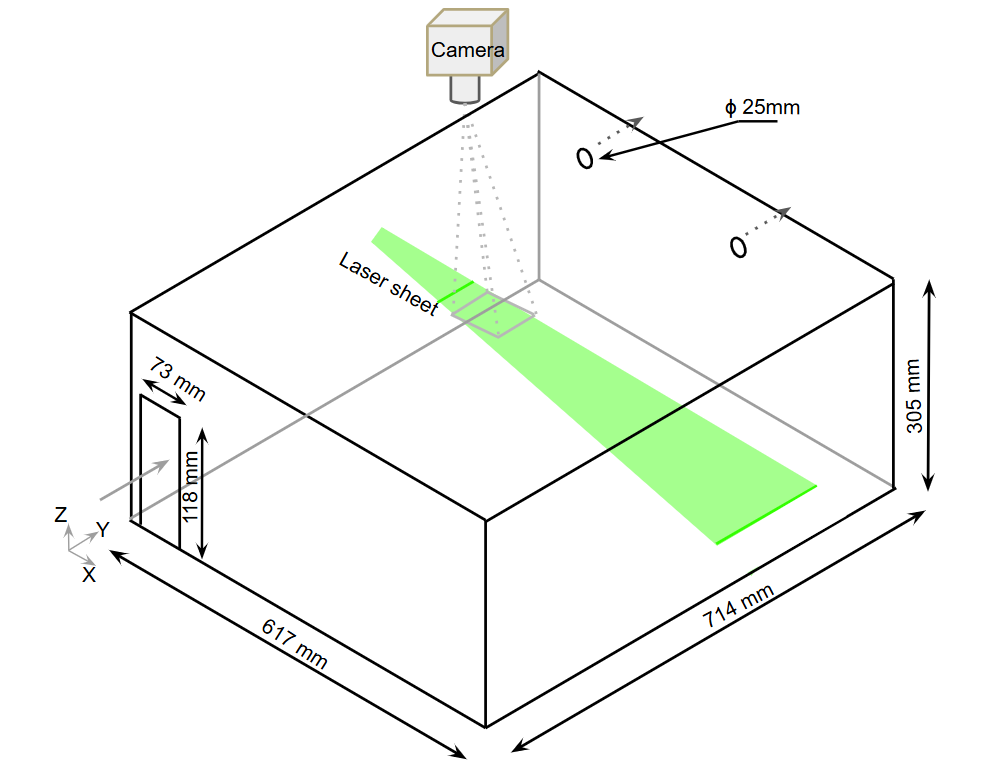} \label{fig:geom2}} \\
\subfloat [] {\includegraphics[width=0.4\textwidth] {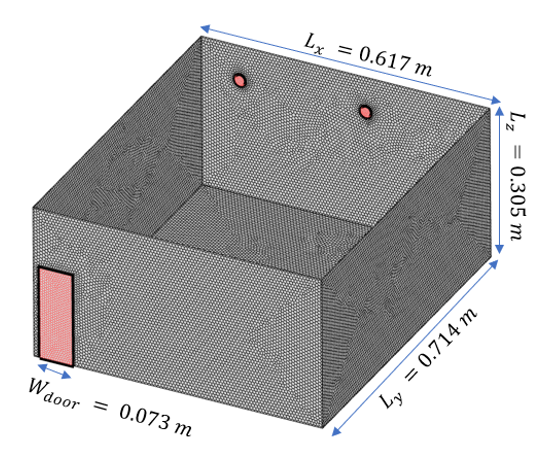} \label{fig:geom3}} 
\caption{Geometry and dimensions of (a) realistic classroom, (b) reduced-scale model, showing PIV set up, and (c) computational domain and grid used in CFD. A brief schematic of the PIV set up is also included in part (b).}
\label{fig:geom}
\end{figure}

The classroom geometry in this work is motivated by one of the real-scale classrooms at Indian institute of Technology Bombay. 
The ventilation in the classroom is via four ceiling fans and two exhaust fans. The dimensions of the room and the locations of the ceiling and exhaust fans are given in Fig.~\ref{fig:geom1}. Ceiling fans are installed for occupant comfort during hot weather, while the exhaust fans throw out use/stale air from the room to outside. The primary source of fresh air entering the room is through an open door. The door is located near the board wall, also called the front wall of the classroom. In winter scenario, the ceiling fans are usually switched off, while the exhaust fans are operational to draw in fresh air and dilute bio-aerosol concentration in the room. 
  
The laboratory model, also mentioned here as "benchtop" model, is a scaled down version ($1/12$ th of original dimensions) of the original classroom; see Fig.~\ref{fig:geom2}. To maintain geometric similarity, the exhaust fan openings are also scaled down by a factor of 12. The volume flow rate and speed of the fan are adjusted to get comparable Reynolds number. This requires a conical duct to match the exit diameter on the exhaust wall of the model and the size of the miniaturized exhaust fans. The current study represents the winter-scenario where ceiling fans are not included in the benchtop model.

The computational domain matches the dimensions of the benchtop model and is shown in Fig.\ref{fig:geom3}. It consists of the door as an inlet and the exhaust ports are modeled as outlet boundary. The blades of exhaust fans are computationally expensive to simulate. They are installed at the end of the conical duct, which is not part of the computational domain. It is expected that the fan blades will not affect the airflow in the fluid domain significantly. Therefore, the effect of the exhaust fan blades is neglected and the exit ports are modeled as velocity outlet. The measured values of the air flow velocity at the exit ports (inside the conical duct) is reported to be 12 m/s; this is specified at the exit boundaries.

\begin{figure*}[t!]  
{\includegraphics[width=0.90\textwidth]{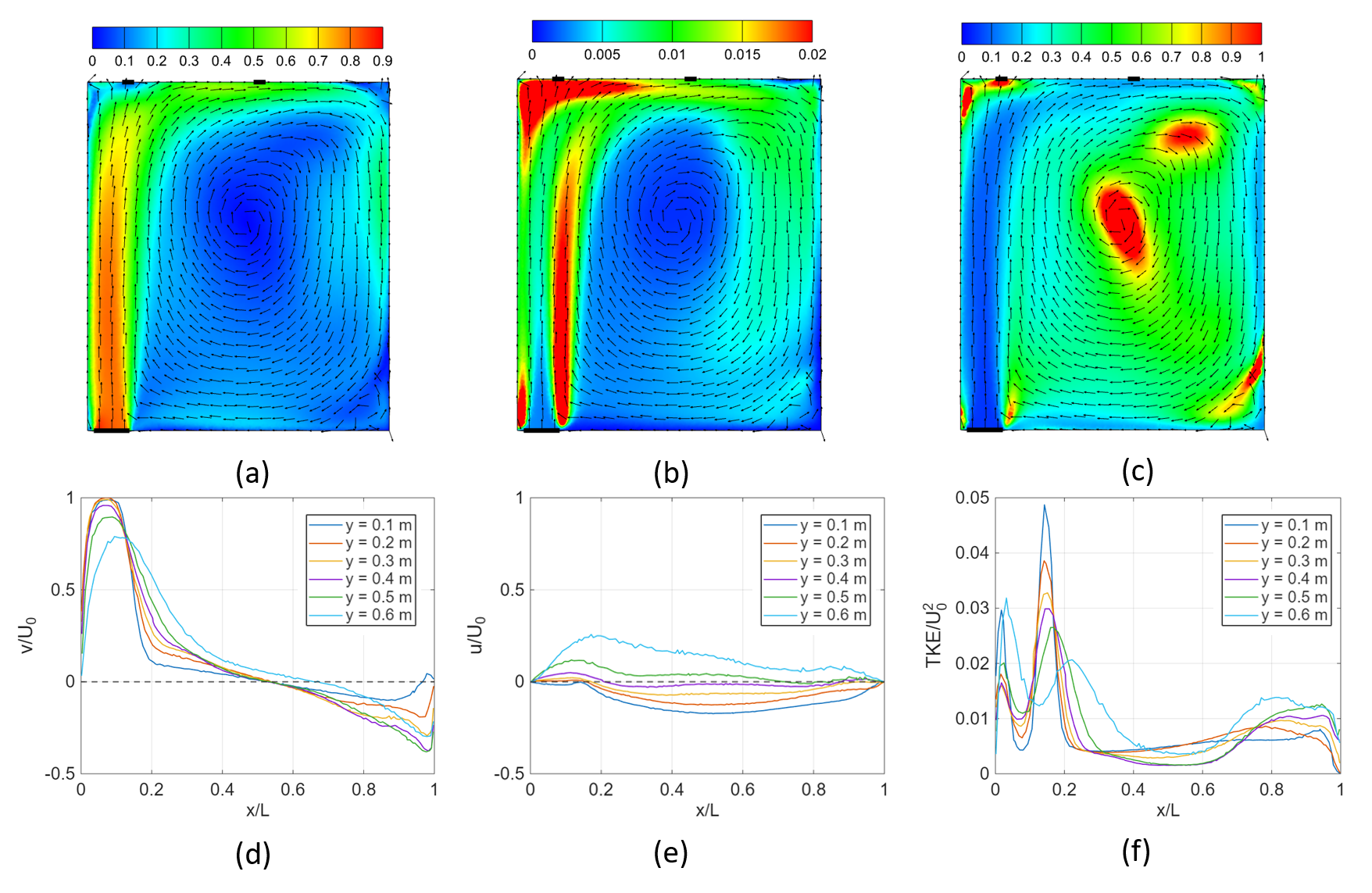}} 
\caption{Flow caracteristics in the scaled breathing level at z$=0.11$ m: (a) velocity magnitude, (b) turbulent kinetic energy and (c) turbulent intensity, along with the profiles of (d) $y$-component and (e) $x$-component of velocity, and (f) turbulent kinetic energy at different  $y$-locations.
}
\label{fig:topview}
\end{figure*}

\subsection{Experimental details}

Air velocity in the scaled-down classroom model was measured using high-speed particle image velocimetry (TSI). The image plane was illuminated by a dual pulsed laser (527 nm, Nd:YAG, Model YLF50-1000-NG) at 250 Hz. Flow was seeded with an aerosol generator (TSI 9307). Images were captured using a CMOS camera (Phantom Veo 710L with 50mm lens) operated at 500 fps and with a spatial resolution of 4 px/mm. The camera and laser were synchronized using TSI LaserPulse 610036. Images were captured at the appropriately scaled breathing height (z=0.11 m) in the model geometry. The breathing plane ($x-y$ at $\frac{z}{H} = 0.36$) was divided into 18 blocks, and instantaneous velocity was measured in each of them. For each experiment, images were captured for 2 seconds. Experiments were repeated at least twice for each box.

Image processing was performed using PIVLab version 3.10. First, a pixel-wise mean intensity subtraction algorithm removed the background interference. Application of Contrast Limited Adaptive Histogram Equalization (CLAHE) enhanced visibility and improved vector accuracy. The Fast Fourier Transform (FFT) window cross-correlation algorithm was employed to calculate velocity vectors. A multi-pass processing approach refined the vector field, starting with a 128-pixel interrogation window and progressively decreasing to 64 and 32 pixels with 50\% overlap to obtain velocity at a 4 mm spatial resolution. Point velocity of air was obtained by averaging the instantaneous velocity over 1500 realizations in each block. At the overlapping boundaries of the blocks, the velocities were spatially averaged to obtain a smooth variation in the velocity field.
In addition, local mean velocity magnitudes were also measured using a hot wire anemometer (Lutron AM-4224SD, with a resolution of 0.01m/s). The velocity obtained using two different methods is compared.

\if{false}

Air velocity in the scaled-down classroom model was measured using high-speed particle image velocimetry (TSI). The image plane was illuminated by a dual pulsed laser (527 nm, Nd:YAG, Model YLF50-1000-NG) at 250 Hz. {\bf Flow was seeded with aerosol generated (oil seeder- TSI 9307), Images were captured using CMOS camera (Phantom Veo 710L with 50mm lens) operated at 500 fps. Camera and laser were synchonized with (TSI LaserPulse 610036). }Images were captured at the scaled breathing plane (z =  0.11m, z/H = 0.36 ). The xy plane was divided into 18 blocks, and velocity was measured in each of them. Images were captured for 2 second and 500 sets of images were captured for each box with three technical replicates. 

Images were processed using PIVLab version 3.10. In the preprocessing, to eliminate static background features, a pixel-wise mean intensity subtraction was first applied across the image set. CLAHE (Contrast Limited Adaptive Histogram Equalization), the default filter in PIVlab, was employed to locally enhance contrast, thereby improving the visibility of flow structures and enhancing vector accuracy. Further, the velocity fields were obtained using the FFT window cross-correlation algorithm, which was found to yield consistent results across different correlation methods.

 A multi-pass processing approach was adopted, beginning with a 128-pixel interrogation window, followed by 64-pixel and 32-pixel windows in successive passes with 50\% overlap. This iterative refinement enhanced vector resolution and signal-to-noise ratio. The same processing settings were applied to all datasets to maintain consistency and reliability. Instantaneous velocities at any point (x,y) was averaged over time.

Further, velocities were averaged over the y-length of the box and the x-variation at the centre of the box is plotted (link with figure). Velocity components in the x-z plane at y = 0.05m were measured to check the relative magnitude of the z-component of the velocity in the breathing plane and establish the accuracy of the planar velocity measurement. The z-component of the velocity was found to be atleast $\mathcal{O}(10^{-1})$ smaller compared to the other two components.  

In addition, local mean velocity magnitudes were also measured using a hot wire anemometer (Lutron AM-4224SD, resolution of 0.01m/s for a range of 0.01 to 5m/s and 0.1m/s for a range of 5 to 25m/s ). The velocity obtained using the two different methods is compared.

\fi

\subsection{Computational grid}

The computational grid is generated using ANSYS Fluent 2022. Geometry was prepared using SpaceClaim. An unstructured polyhedral mesh with $3.56 \times 10^6$ elements is used as a baseline grid; see Fig.~\ref{fig:geom3}. The minimum orthogonal quality of $0.37$ in the current mesh is above the commonly recommended threshold of $0.2$, indicating acceptable cell alignment for numerical stability. Similarly, the maximum aspect ratio of $6.5$ falls within the acceptable range (typically $\le 10$) for general CFD applications, as supported by Fluent guidelines.

The first layer thickness at the walls is kept at an average \( y^+ \) value of $3.5$, which indicates that we have refined the near wall region well. Enhanced wall treatment method, suitable for $y^+<5$, is used for applying boundary conditions at the wall for turbulence variables in the k-$\epsilon$ framework.
A mesh independence study is presented in Appendix A, which compares results obtained using a coarser and a finer mesh. The data confirms that the air flow solution is not sensitive to further grid refinement. The baseline mesh is selected based on a trade-off between accuracy and computational cost.

\begin{figure*}[t!]
  \centering  {\includegraphics[width=0.8\textwidth]{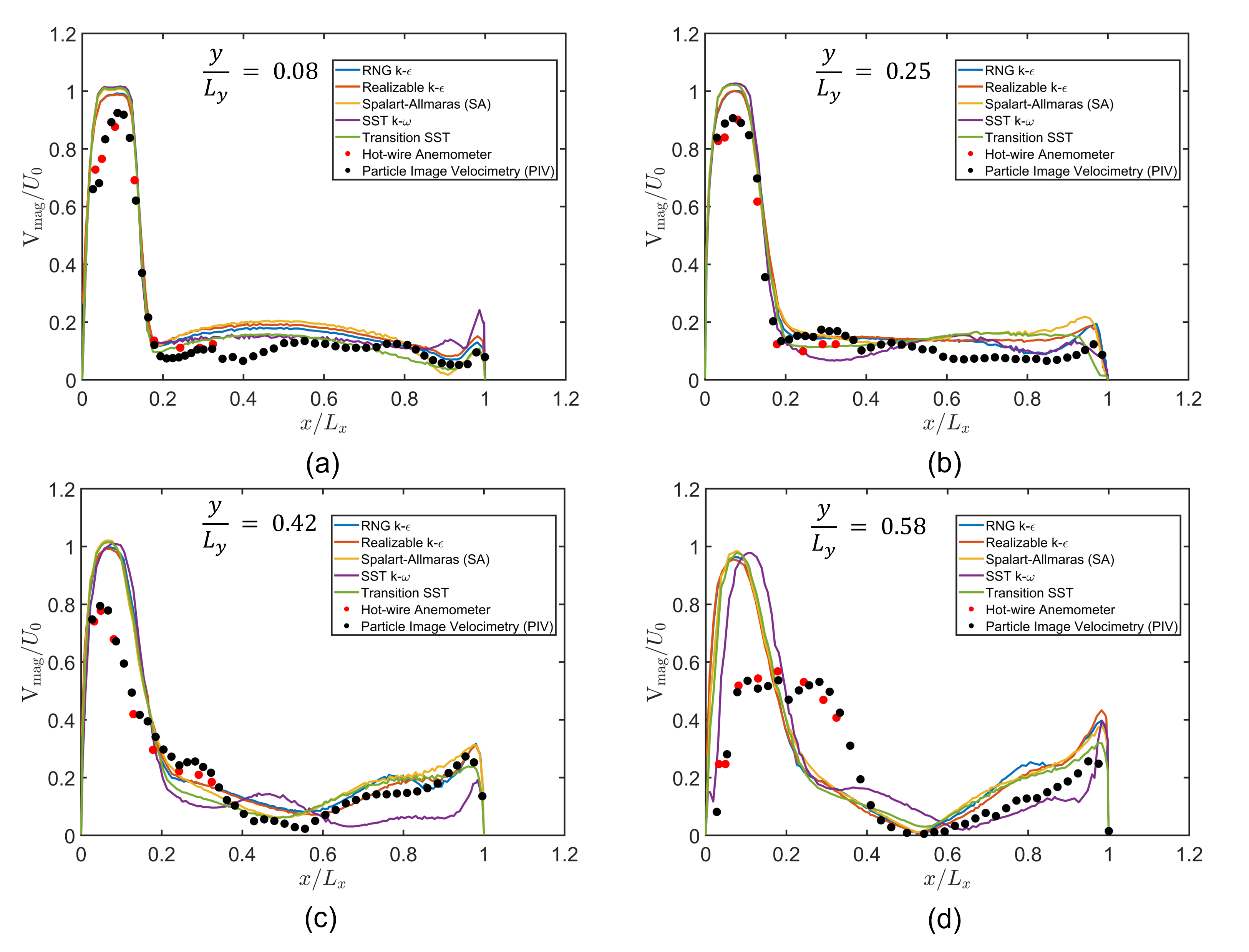}}
\caption{Velocity profile comparison with PIV and anemometer results for different turbulence models at four streamwise $y$-locations.}
\label{fig:velprof_turb}
\end{figure*}

\section{\bf{Results}}

\subsection{Flow characteristics}

Figure \ref{fig:topview} presents the mean flow and turbulent characteristics in the classroom model on a horizontal plane. Air from door entry forms a jet like pattern along the front wall. There is a turning effect at the exhaust fan wall and reversed flow along the back wall. A large recirculation zone forms in the classroom, with relatively low velocity. 
The flow pattern is similar to that in the forced convection test case presented in section II; see Fig.~\ref{fig:validation}. The velocity contours plotted in a vertical plane in Nielsen's case are qualitatively similar to those in a horizontal plane in the current study.
Turbulent kinetic energy in Fig.~\ref{fig:topview}b has high values in the shear layer enclosing the jet flow along the wall. There are large TKE values near the exhaust fan wall and relatively low TKE in the center of recirculation flow. Turbulence intensity plotted in Fig.~\ref{fig:topview}c in turn shows high values in the recirculation region, indicating that fluctuations are large compared to the mean flow in this region. 

Velocity profiles plotted along constant y-lines (Fig.~\ref{fig:topview}d) show the streamwise development of the jet flow. Peak velocity in the $y$-direction decreases with distance, while the spread increases in the $x$-direction. High negative velocity is observed in the reversed flow at the back wall. 
Fig.~\ref{fig:topview}e shows that cross-flow $x$-velocity has a relatively low contribution to the overall velocity magnitude. The maximum $x$-velocity is within $\pm 0.25$ of the peak streamwise velocity.
TKE profiles in Fig.~\ref{fig:topview}f show the prominent peak in the shear layer ($x/L \approx  0.15$ m) and a second peak near the front wall ($x = 0$). 

\subsection{Different Turbulence models}

Figure \ref{fig:velprof_turb} shows the velocity profile obtained from different turbulence models, and their comparison with experimental measurements. The data is plotted at four streamwise $y$-locations, at $z$ = 0.11 m, which corresponds to the scaled-down breathing plane in the model classroom. The streamwiwe $y$ and lateral $x$ coordinates are non-dimensionalized by the geometric length in the respective directions; see Fig.~\ref{fig:geom}. The magnitude of air velocity $V_{mag}$ is normalized by the nominal average inlet velocity at the door ($U_0$).
There are two important points to note from the figure. 

First, there is negligible difference between the model predictions in the high-velocity jet flow ($x/L_x < 0.2$). This is similar to the forced convection case presented in section II.
The low velocity recirculation region shows relatively larger variability between the turbulence models.
The SST $k-\omega$ model seems to be an outlier, with larger differences with the measurements in the recirculation region; see parts (c) and (d) of the figure. 


%
%


\if{false}
\begin{figure*}[t!]  
{\includegraphics[width=0.8\textwidth]{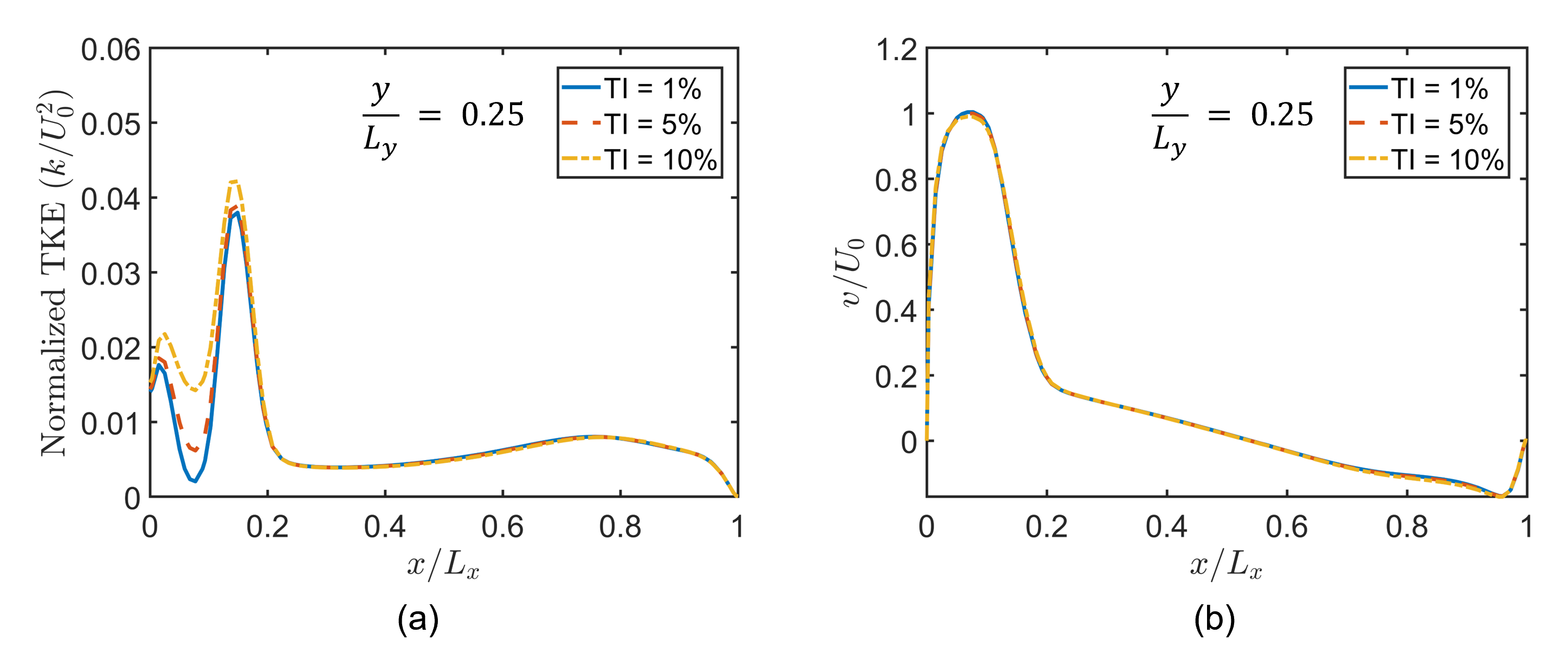}} 
\caption{Effect of varying turbulence Intensity at the inlet boundary on velocity profile and the distribution of TKE at two $y$ locations.}
\label{fig:inlet_ti}
\end{figure*}
\fi

\begin{figure*}[t!]  
\subfloat []{\includegraphics[width=0.4\textwidth]{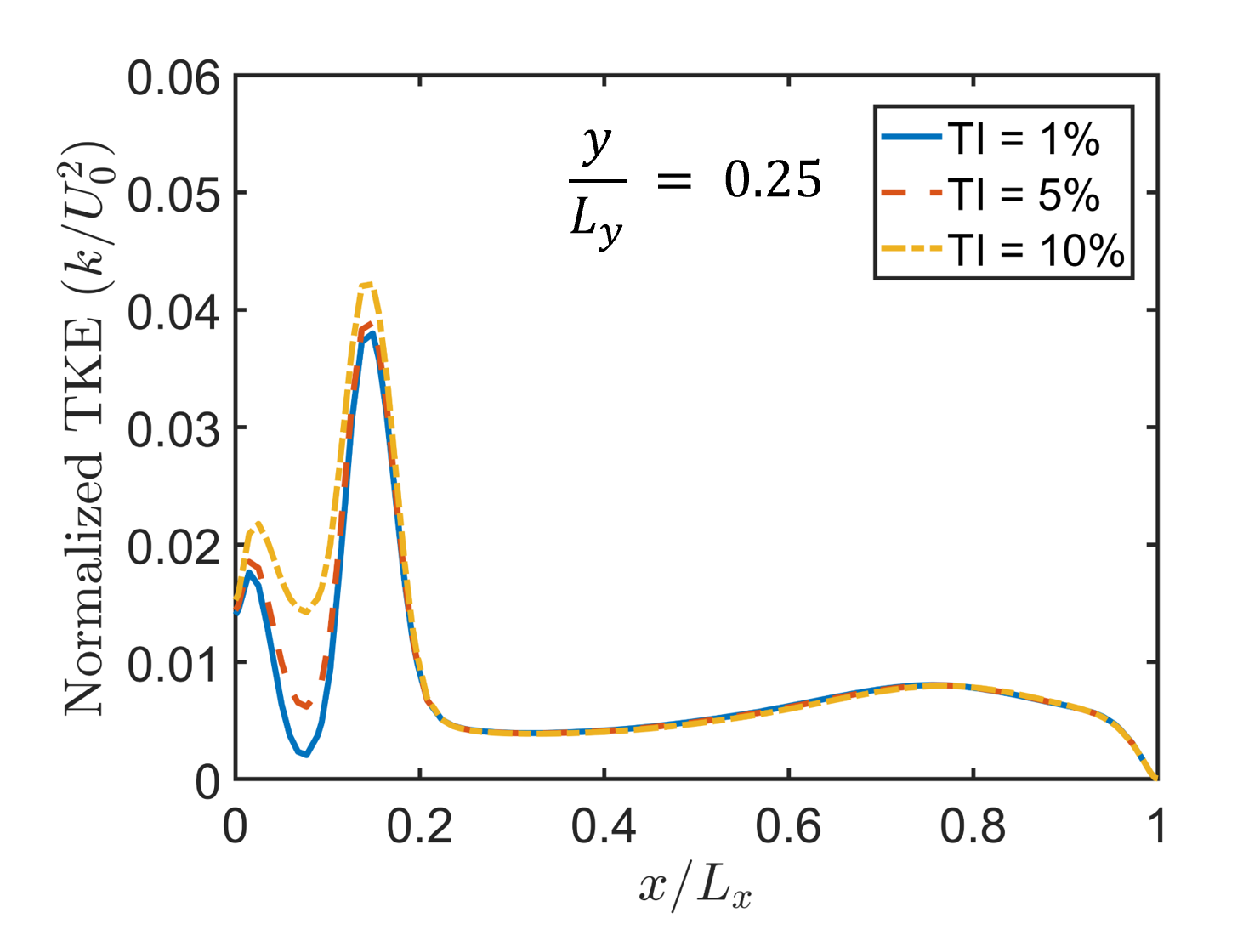} \label{fig:Fig5_1}} 
\subfloat [] {\includegraphics[width=0.4\textwidth]{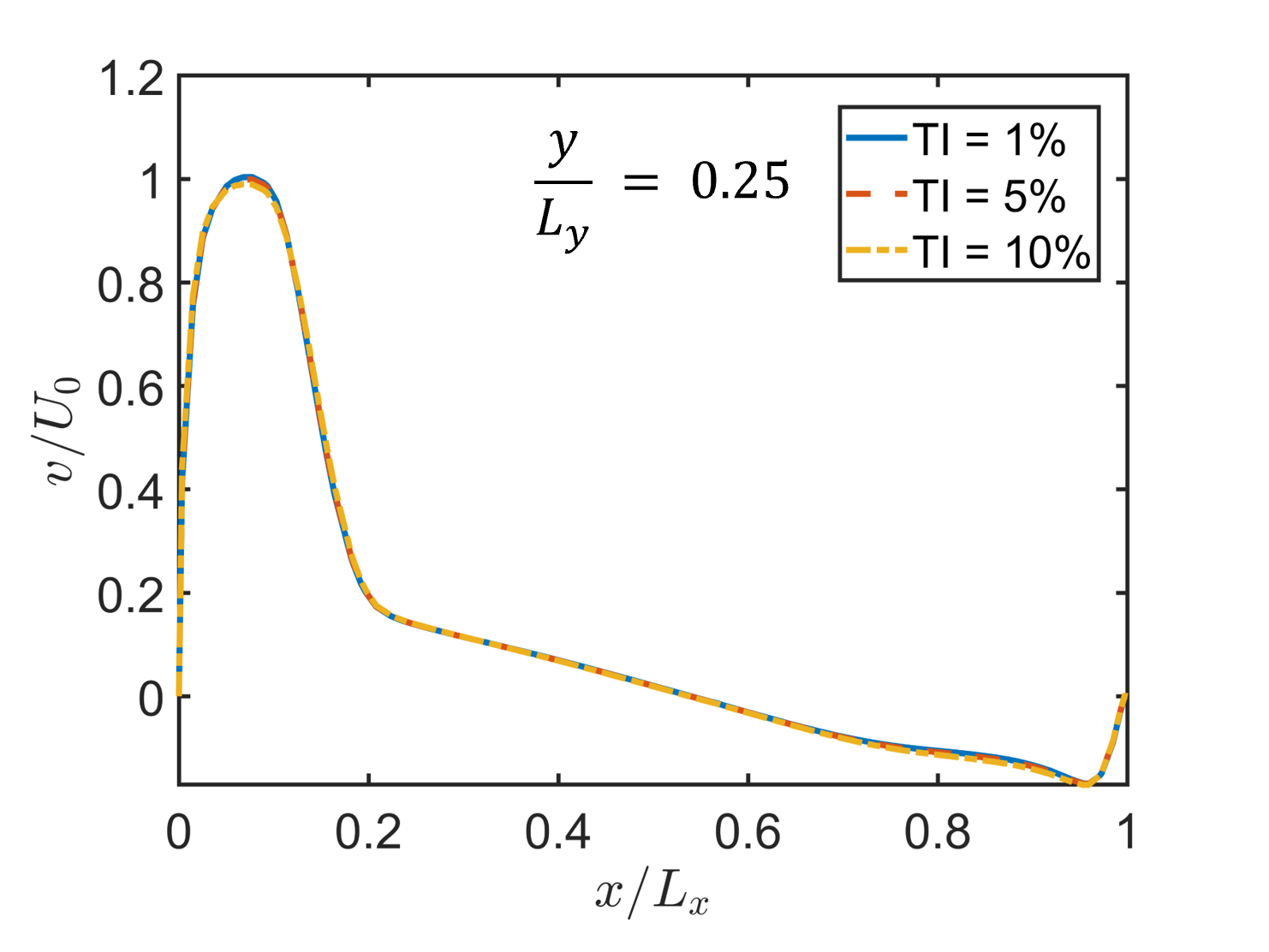} \label{fig:Fig5_2}} 
\caption{Effect of varying turbulence Intensity at the inlet boundary on (a) TKE, (b) velocity}
\label{fig:inlet_ti}
\end{figure*}


Second, comparison between CFD and experiments show that 
the turbulence model predictions are consistently higher than the measurements. The discrepancy is small in magnitude at lower values of $y$, i.e. closer to the inlet door (see Fig.~\ref{fig:velprof_turb}(a) and (b)), while it increases dramatically further downstream (higher $y$ locations). Thus peak velocity decays much faster in the experiment than that predicted by CFD. The spreading of the shear layer is also found to be higher in the experiment, as evident in part (c), and a broader peak in part (d) of the figure.

The recirculation region has low velocity magnitude, and the values predicted by CFD are comparable to the measurements. The local high velocity at x = 0.6 at line 4 indicates a strong reversed flow. This is similar to that observed in Nielsen's test case, except for a change in sign (y-velocity vs. velocity magnitude). The CFD velocity profiles are qualitatively similar to those plotted in Fig.~\ref{fig:validation}c and ~\ref{fig:validation}d and a comparison between the present case with the validation test case with appropriate normalization and scaling is presented below. This is found to be critical to explain the observed differences between CFD and experimental measurements.

\begin{figure*}[t!]  
{\includegraphics[width=0.9\textwidth]{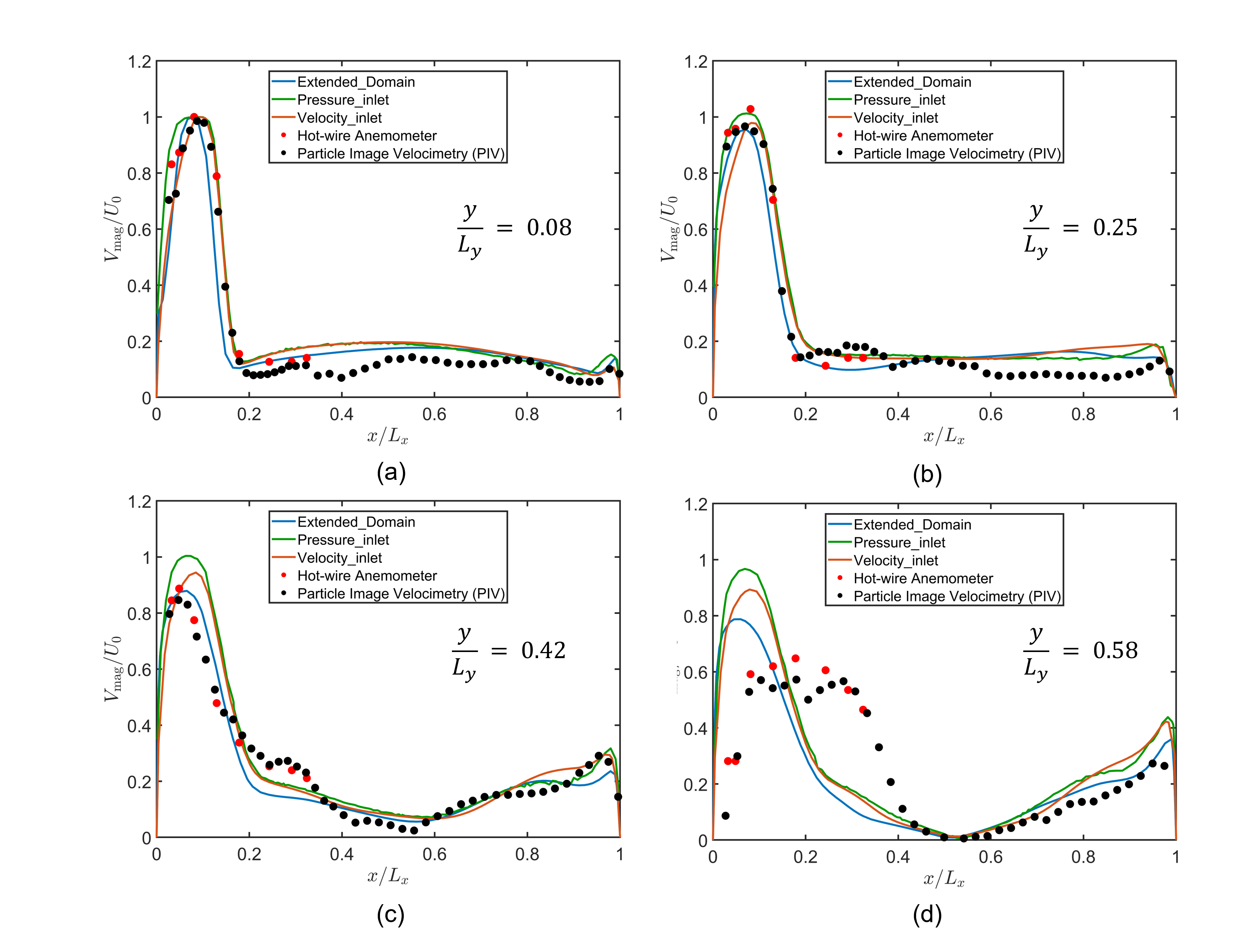}} 
\caption{Effect of inlet boundary conditions on the computed velocity profile, and its comparison with experimental data.}
\label{fig:inlet_bc}
\end{figure*}

\subsection{Effect of boundary condition}

The outdoor air entering the domain through the open door is similar to a natural ventilation scenario. 
Outdoor conditions can have an effect on the mean and turbulence velocities at the door.
These can affect the boundary conditions specified at the door entry.
Here, we study the effect of variations in mean flow boundary condition and turbulent intensity TI specified at the door on the airflow characteristics inside the domain.

Figure \ref{fig:inlet_ti} shows the profiles of turbulent kinetic energy and mean $y$-velocity for three different values of inlet turbulent intensities. The baseline intensity of 5\% is compared with a low intensity of 1\% and a high intensity of 10\% specified at the door. As expected, the turbulent kinetic energy increases with increase in TI, but there is no noticeable change in the mean flow velocity. The trend is similar at other $y$-locations, not shown in the figure. 

Next, we study the effect of pressure vs. velocity boundary condition at the door entry and exit locations. The baseline case presented above has specified pressure with zero gauge value at the door entry and a specified exit velocity condition at the exhaust fan boundary. We now reverse the boundary conditions by specifying the velocity at the door (as per measurements available from the experiment) and make the exit locations as pressure outlet. 
The velocity profiles for the different boundary conditions are compared in Fig.~\ref{fig:inlet_bc}. The pressure and velocity boundary condition results are close to each other, with minor differences in the high-velocity primary flow. However, we observe a faster decay of the peak velocity when the door boundary is specified in terms of inlet velocity.

We also compute an extended domain, where the space outside the door is included in the computational domain. Details of extended geometry, computational grid and flow velocity at the door are given in Appendix B. The velocity profiles at the four streamwise locations is included in Fig.~\ref{fig:inlet_bc}. It is found that the extended domain shows a faster decay of the peak velocity compared to the pressure and velocity inlet cases. The extended domain velocity profiles match the experimental data for $y/L$ = 0.08, 0.25 and 0.42. The discrepancy at $y/L$ = 0.58 is still significant. 

We note that the CFD and experimental data presented in Fig.~\ref{fig:inlet_bc} are normalized by their respective maximum velocity at the door $U_{max}$. In case of the extended domain, we use the maximum velocity at the neck of vena contracta (see Fig.\ref{fig:xy vena contracta} in appendix) as the maximum velocity for scaling the velocity profiles in this case.  




\begin{figure*}[t!]  
\subfloat []{\includegraphics[width=0.45\textwidth]{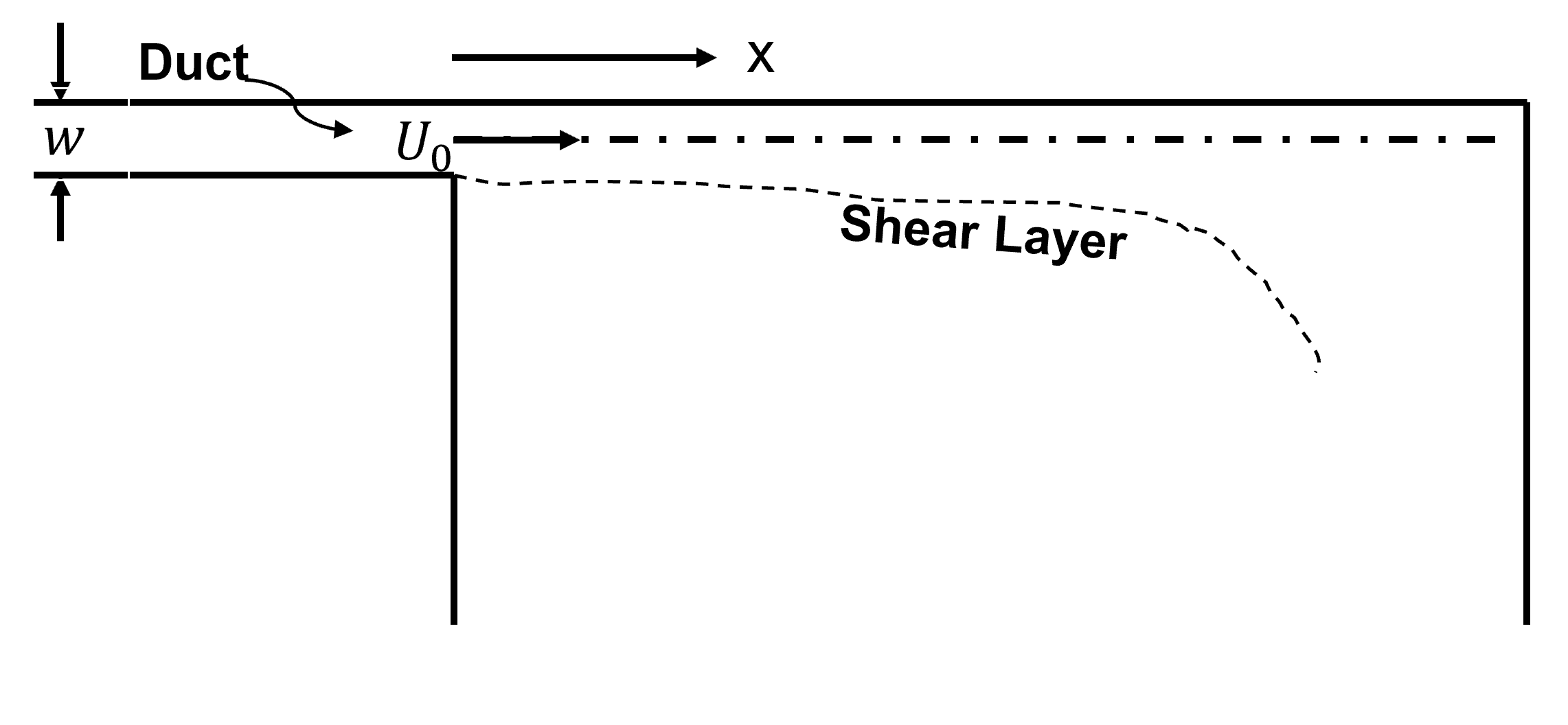} \label{fig:sch1}}
\subfloat [] {\includegraphics[width=0.4\textwidth]{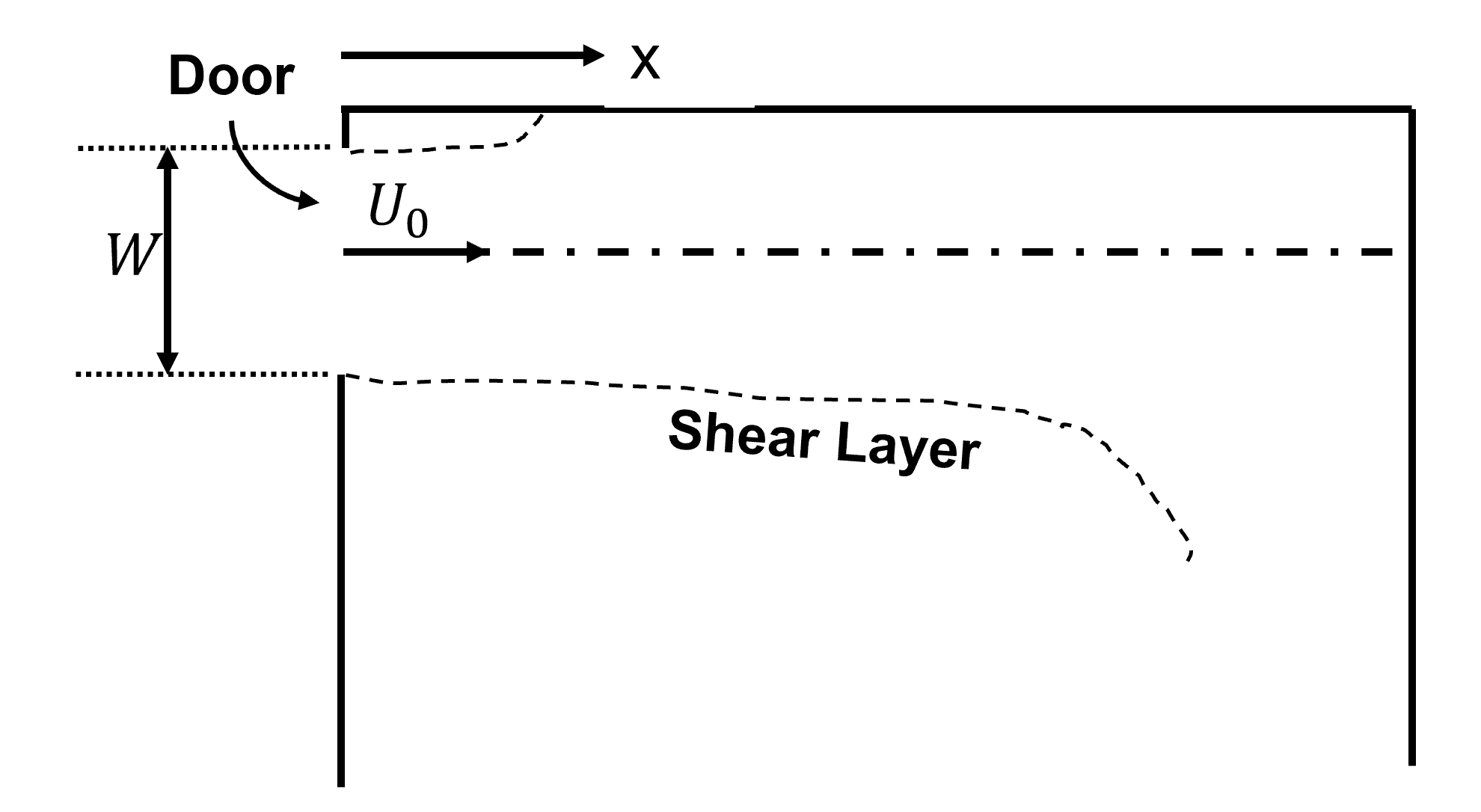} \label{fig:sch2}} 
\caption{Schematic of (a) Nielsen's experiment and (b) reduced-scale classroom model to study the scaling of the inlet jet centerline velocity decay.}
\label{fig:schem}
\end{figure*}

\section{Centerline velocity scaling}

Experimental and computational data for the reduced-scale classroom model are compared in terms of the jet center-line velocity. The figure also shows the forced convection test case from Section II for reference. Both geometries are schematically shown in Fig.~\ref{fig:schem}, along with their characteristic dimension and velocities. The jet shear layers and the centerline are marked to show that the flow in the two cases are expected to be qualitatively similar. 

\begin{figure}[t!]
  \centering
    \includegraphics[width=0.5\textwidth]{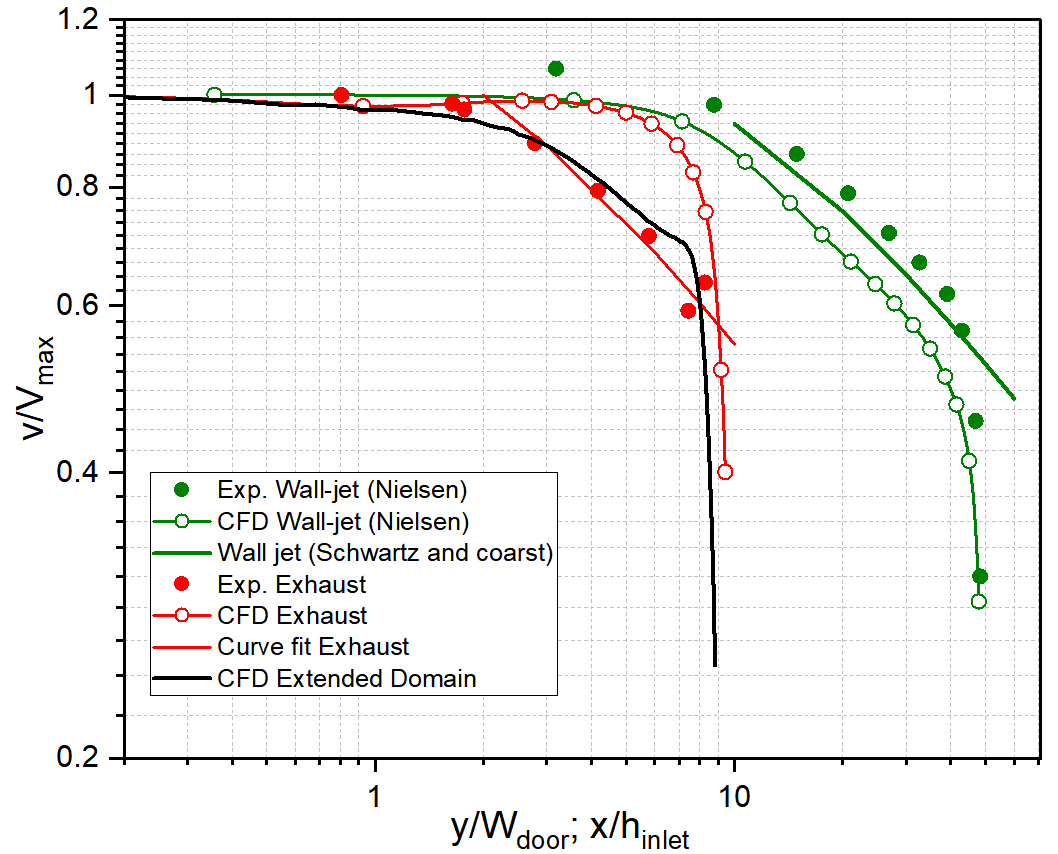} \\ 
\caption{Centerline velocity plotted as a funciton of distance along the jet centerline on log-log scale to compare computational, experimental and analytical solutions.}
\label{fig:centerline}
\end{figure}

For the forced-convection test case of Nielsen,\cite{nielsen1978velocity} the centerline velocity is normalized by the jet entry velocity (the maximum value at the inlet), and the distance from the inlet is normalized by the width $w$ of the inlet slot. Our data  matches the results presented by Nielsen;\cite{nielsen1978velocity} see Fig.~\ref{fig:centerline}.
The normalized velocity is close to unity for $x/w < 10$, which corresponds to the potential core. This is followed by a power law decay that appears as a linear part in the log-log plot in the figure. There is a rapid drop in the velocity as the flow reaches the end wall ($y/w \sim 50$).

Both experimental \cite{nielsen1978velocity} and our CFD results for Nielsen's case match the slope of the analytical wall-jet scaling of Schwartz and Coarst.\cite{schwarz1961two}
\begin{equation}
\frac{V_m}{V_0}=C_u\left(\frac{x-x_D+x_0}{h}\right)^e
\end{equation}

\noindent
where $V_m$ is the maximum velocity at a given streamwise location $x$ and $V_0$ is a reference velocity, typically taken as the maximum velocity at the inlet. There is very little difference between the maximum velocity and the centerline velocity plotted in Fig.~\ref{fig:centerline}. A characteristic dimension of the inlet is given by $h$, whereas $x_D$ and $-x_0$ are the coordinates of the inlet plane and the virtual origin of the jet. The values of the parameters are $C_u = 5.5$, $e = -0.55$, $x_D = 0$ and $x_0/h = 11.2$ are taken from the reference.\cite{nielsen1978velocity}

The experimental data for the current classroom ase (red filled symbols) shows a qualitatively similar trend. The normalized velocity approaches one at lower limit of $x/W$ and shows a rapid drop at the end ($x/W \sim 8$). The $x/W$ range is substantially smaller in the current case primarily due to the a higher value of the door width $W$ compared to the slot width $w$ used for normalization in Nielsen's case.  Interestingly, the linear decay rate observed in the experimental data (red line) is found to be parallel to the wall-jet scaling of Schwartz and Coarst. It corresponds to Eq.~(1) with $C_u = 2.14$, $e = -0.55$, $x_D = 0$ and $x_0/h = 2$

CFD solutions obtained using two distinct inlet boundary conditions are plotted in terms of their centerline velocity decay. The classroom model case with pressure inlet boundary condition corresponds to the Realizable $k$-$\epsilon$ model and it is representative of the other turbulence models used in the CFD simulations. The center-line data seem to follow the wall-jet CFD data obtained for the forced convection case of mechanical ventilation, up to about $x/W = 5$. The CFD data falls off beyond this point, leading to a rapid drop due to end wall effect.
The CFD simulation with velocity inlet boundary condition is  close, but slightly lower than the pressure inlet solution. It is not shown in the figure for the sake of clarity.

On the other hand, the CFD results obtained from the extended domain simulation is found to align well with the experimental data for the reduced-scale classroom model. The centerline velocity in this case is normalized by the maximum velocity at the vena contracta at a distance of 0.05 m from the door (see Fig.~\ref{fig:xy vena contracta} in Appendix). This is assumed to be the virtual origin of the jet, and the $x/W$ values are accordingly calculated. 

The data presented in Fig.~\ref{fig:centerline} brings out two important points.
First, the drop in centerline velocity is much faster in the classroom model than in the forced convection test case. 
We note that the inlet jet in the current case is driven by the suction effect of the exhaust fans, as opposed to the forced jets in Nielsen's mechanical ventilation test case. 
This indicates that there may exist a different scaling for the velocity decay in hybrid ventilation.
Second, the plot brings out the strong influence of the inlet boundary condition on the jet flow in the domain. It points to the fact that a detailed characterization of the inlet flow may be required for such cases of hybrid ventilation. This is similar to what is observed for natural ventilation, where indoor airflow is sensitive to geometry of the building and openings like doors and windows. Future work in this direction is required to study the effect of external geometry and air flow in terms of mean and fluctuating velocity measurements outside the door.


\section{Conclusion}

In the present work, we study the airflow in a reduced-scale classroom model with hybrid or mechanically-assisted natural ventilation. The airflow is driven by a pair of exhaust fans that throw out used air from the room; fresh outdoor air enters through an open door.
CFD simulations are are performed using different RANS based turbulence models, and different inlet boiundary conditions. Results are compared with detailed velocity profiles measured using particle image velocimetry.

A good agreement between the CFD and experimental results is observed close to the door entry. However, the agreement is poor away from the door, at locations that are closer to the suction generated by the exhaust fans. Most notably, the peak velocity in the inlet jet is much lower in the experiment, indicating higher level of mixing and spread of the jet shear layer.
Very little variation was observed between the different turbulence models, including low Reynolds number transitional models.

The CFD results are found to be sensitive to inlet boundary conditions in the form of prescribed pressure versus prescribed velocity values. The largest variation are observed when the computational domain is extended to include the space outside the door. The extended domain results are closest to the experimental data. 
In particular, the centerline velocity in the jet flow entering the door aligns with the PIV data, when normalized by the maximum velocity near the door. On the other hand, the centerline velocity data from the other CFD simulations (prescribed pressure or prescribed velocity at inlet boundary) align with existing correlation for wall jets.

Overall, the paper presents a detailed comparison of CFD and experiments for hybrid ventilation airflow in a scaled-down model of a realistic classroom. It is a unique contribution to the existing literature of CFD validation for indoor air flow simulations.

\if{false}   
Additional experiments and simulations were performed to explain the observed discrepancies, and centerline velocity data are compared with known scaling relations. 
It was found that CFD and experiments match Pope's scaling in case of forced jet, without door and exhaust. Similar match is also shown for Nielsen's mechanical ventilaton test case.
By comparison, experimental data for hybrid ventilation with exhaust and open door show new trends for centerline velocity scaling.
CFD results seem to conform to well-established scaling, and therefore deviate from experimental data.


On the whole, the observations of the current work for hybrid ventilation is consistent with earlier works that report poor performance of CFD models in predicting natural ventilation.
This work points to the need for further research in this area.


\fi      


\section*{References}
\bibliographystyle{unsrtnat}
\bibliography{Mybib}




\section*{Appendix}
\subsection{Grid Sensitivity}

\begin{figure}[t!]
  \centering
  \subfloat []{\includegraphics[width=0.4\textwidth]{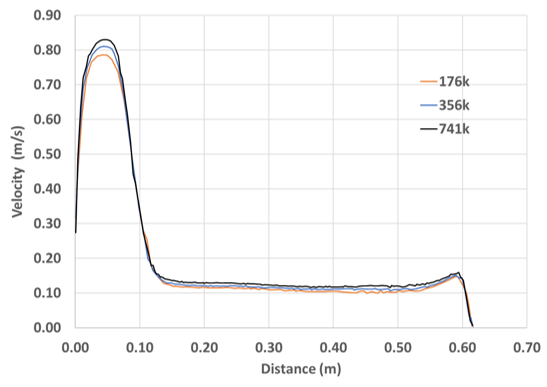}} \\
  \subfloat []{\includegraphics[width=0.4\textwidth]{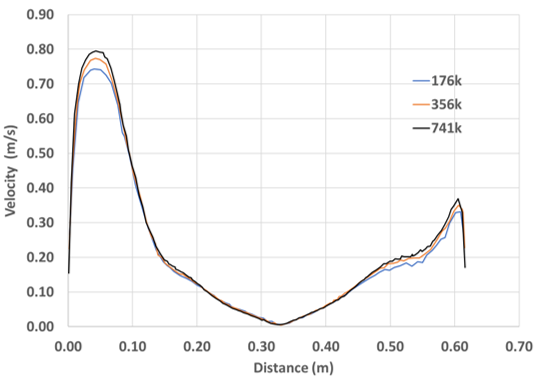}} 
\caption{Velocity profiles along $y=0.1785$ m and $y=0.4165$ m lines computed using different grid resolutions.}
\label{fig:vel_grid_conv}
\end{figure}

The sensitivity of the results to the computational grid is evaluated by comparing the solutions obtained on successively refined grids. The baseline grid of 3.56$\times 10^5$ is complemented by a coarser
grid of 1.76$\times 10^5$ and a finer grid of 7.41$\times 10^5$ elements. The magnitude of velocity is plotted along horizontal lines spanning the entire room from the board wall to the back wall at constant $y$ values of $0.01785$ m and $0.4165$ m respectively. The data is presented in Fig.~\ref{fig:vel_grid_conv} as a function of the distance from the board wall; see section III. All three velocity profiles are qualitatively similar; they show a region of high velocity within 0.01 - 0.07 m of the board wall, a region of low velocity in the middle of the classroom (0.15 - 0.5 m), and relatively higher velocity near the back wall (0.55 - 0.6 m).
More important, we note a good match between the results computed on the different grids. The peak velocity varies less than 3\% between the coarse and baseline grids, while the fine grid maximum velocity overlaps with the baseline results.

\begin{figure*}[t!]
  \centering
  {\includegraphics[width=0.8\textwidth]{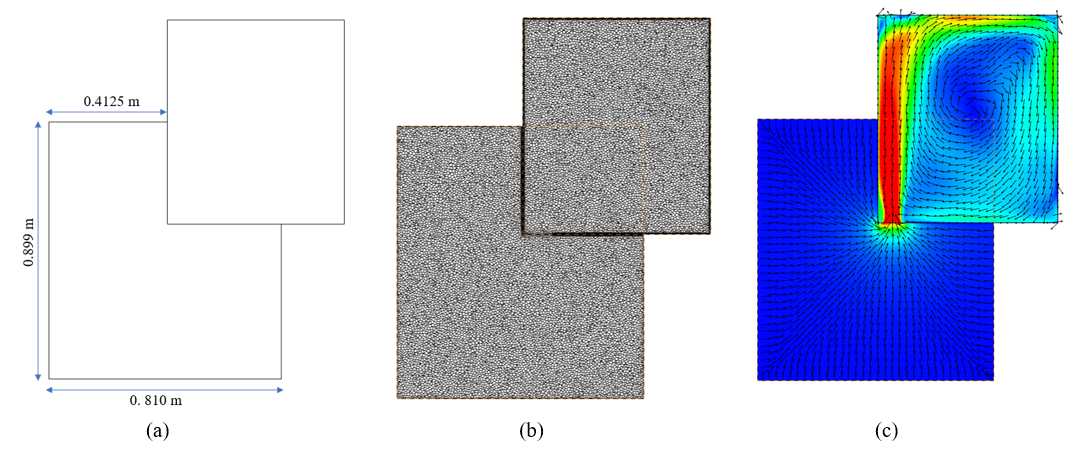}}
\caption{Geometrical dimensions along with computational grid and velocity contour plotted at $z=0.11$m.}
\label{fig:Ext Domain}
\end{figure*}

\begin{figure*}[t!]
  \centering
  {\includegraphics[width=0.9\textwidth]{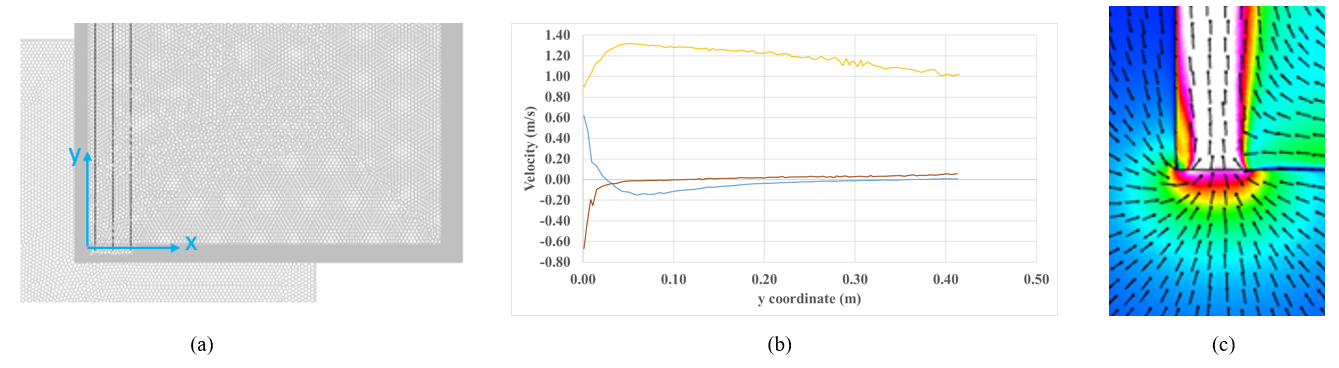}} 
\caption{Representation of xy vena contract at breathing plane $z=0.11$m a) graphical representation of lines plotted at $x=0.02$ m, $x=0.05$ m and $x=0.08$ m (b) y component of Velocity plotted along these lines (c)  }
\label{fig:xy vena contracta}
\end{figure*}

\begin{figure*}[t!]
  \centering
  {\includegraphics[width=0.9\textwidth]{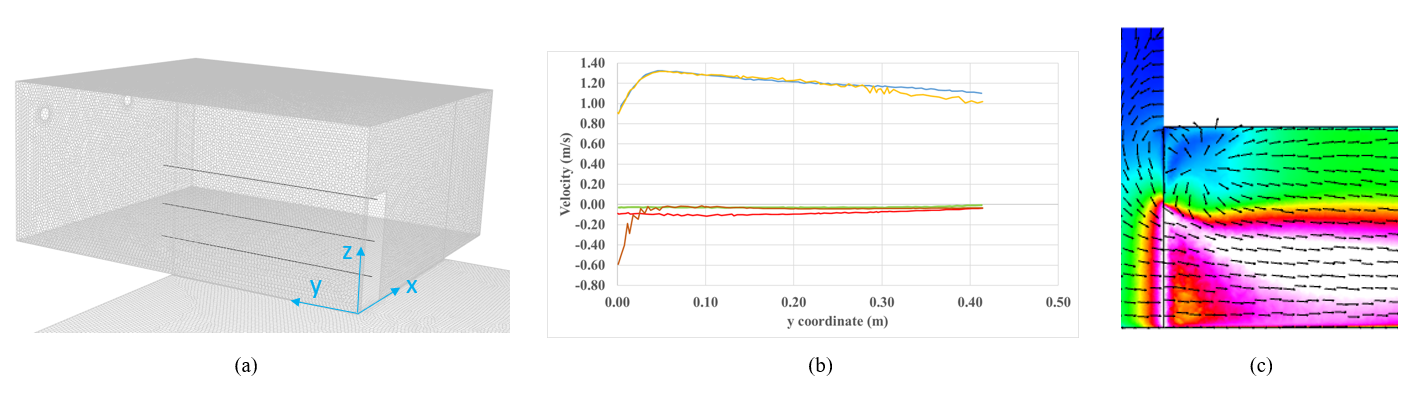}}
\caption{Representation of yz vena contract at breathing plane $z=0.11$ m a) graphical representation of lines plotted at $x=0.02$ m, $x=0.05$ m and $x=0.08$ m (b) y component of velocity plotted along these line}
\label{fig:yz vena contracta}
\end{figure*}

\subsection{Extended domain simulation}

The CFD results presented in Section III.B are obtained by setting the door as a pressure inlet boundary and the velocity is determined by the exhaust fan volume flow rate along with the door area. 
An alternate configuration was simulated, where the space outside the door is also included in the computational domain, with zero guage pressure set at the boundaries of the extended domain. This is to minimize the effect of boundary condition at the entry of the jet flow through the door. 
The geometric details, the computational grid and the mean velocity field are shown in Fig.~\ref{fig:Ext Domain}.
The flow field is similar to that presented in Fig.~\ref{fig:topview}, with a well defined jet like flow along the front wall of the classroom model, flow turning at the exhaust fan wall and a reversed flow along the back wall. A large low-velocity recirculation zone forms in the middle of the classroom space.

A magnified view of the door entry location in Fig.~\ref{fig:xy vena contracta} shows flow entrainment from all sides to form a flow structure similar to vena contracta at the door. The mean velocity components are plotted along three lines shown in part (a) of the figure. The centerline $y$-velocity shows a peak at about 0.05 m followed by a decay in the streamwise direction. By comparison, the $x$-velocity plotted at two edges of the door show large positive and negative values respectively. This indicates three-dimensional flow entrainment, leading to a large axial velocity in the door jet. The peak value is close to 1.3 m/s, as compared to 0.8 m/s in Fig.~\ref{fig:topview}.

A vena contracta is also visible in the $z$-direction (see Fig.~\ref{fig:yz vena contracta}), where the flow is entrained from above the door. The $z$-velocity plotted near the top edge of the door shows large negative values, while the $x$-component is comparatively small.  Overall, simulating an extended domain results in large changes in the velocity field at the door entry, with significant three-dimensional effects. Such effects are not present in the simulations presented above (see Fig.~\ref{fig:topview}), where the door is treated as pressure inlet boundary, and the inlet velocity assumes a near uniform value of 0.8 m/s.

\clearpage


\end{document}